\RequirePackage{snapshot}
\documentclass[sigconf]{acmart}
\usepackage{todonotes} %
\usepackage{xspace}
\usepackage{enumitem,kantlipsum}
\usepackage[most]{tcolorbox}
\usepackage{cleveref}
\usepackage{subcaption}
\usepackage{balance}
\usepackage[ruled]{algorithm2e}
\presetkeys{todonotes}{fancyline, color=olive!30, inline}{}
\newcommand{\new}[1]{#1}%
\newcommand{\todoog}[1]{}%
\newcommand{\todoam}[1]{}%
\newcommand{\FlowDNS}{FlowDNS\xspace}
\newcommand{\eg}{e.g.,\xspace}
\newcommand{\ie}{i.e.,\xspace}
\usepackage[absolute,showboxes]{textpos}

\AtBeginDocument{%
  \providecommand\BibTeX{{%
    \normalfont B\kern-0.5em{\scshape i\kern-0.25em b}\kern-0.8em\TeX}}}

\begin{document}

\setcopyright{acmcopyright}

\acmYear{2022}\copyrightyear{2022}
\acmConference[CoNEXT '22]{The 18th International Conference on emerging Networking EXperiments and Technologies}{December 6--9, 2022}{Roma, Italy}
\acmBooktitle{The 18th International Conference on emerging Networking EXperiments and Technologies (CoNEXT '22), December 6--9, 2022, Roma, Italy}
\acmPrice{15.00}
\acmDOI{10.1145/3555050.3569135}
\acmISBN{978-1-4503-9508-3/22/12}
\title[\FlowDNS: Correlating Netflow and DNS Streams at Scale]{\FlowDNS: Correlating Netflow and DNS Streams at Scale}
%
%
%
%
%
%
\begin{comment}
\author{Aniss Maghsoudlou \bigstar, Oliver Gasser \bigstar, Ingmar Poese \Delta, Anja Feldmann \bigstar}
\email{\bigstar aniss,oliver.gasser,anja@mpi-inf.mpg.de, \Delta ipoese@benocs.com}
\affiliation{%
  \institution{\bigstar Max Planck Institute for Informatics, Saarbrucken, Germany, \Delta BENOCS GmbH}
  \city{\bigstar Saarbrucken, \Delta Berlin}
  \country{Germany}
}
\end{comment}
%
    \author{Aniss Maghsoudlou}
    \email{aniss@mpi-inf.mpg.de}
    \affiliation{%
      \institution{Max Planck Institute for Informatics} 
      \city{Saarbr\"ucken}     
      \country{Germany}
    }
    \author{Oliver Gasser}
    \email{oliver.gasser@mpi-inf.mpg.de}
    \affiliation{%
      \institution{Max Planck Institute for Informatics}
      \city{Saarbr\"ucken}
      \country{Germany}
    }
    \author{Ingmar Poese}
    \email{ipoese@benocs.com}
    \affiliation{%
      \institution{Benocs GmbH}
      \city{Berlin}
      \country{Germany}
    }
    \author{Anja Feldmann}
    \email{anja@mpi-inf.mpg.de}
    \affiliation{%
      \institution{Max Planck Institute for Informatics}
      \city{Saarbr\"ucken}
      \country{Germany}
    }
\renewcommand{\shortauthors}{Maghsoudlou, et al.}

\begin{abstract}
Knowing customer's interests, e.g. which Video-On-Demand (VoD) or Social Network services they are using, helps telecommunication companies with better network planning to enhance the performance exactly where the customer's interests lie, and also offer the customers relevant commercial packages.
However, with the increasing deployment of CDNs by different services, identification, and attribution of the traffic on network-layer information alone becomes a challenge:
If multiple services are using the same CDN provider, they cannot be easily distinguished based on IP prefixes alone.
Therefore, it is crucial to go beyond pure network-layer information for traffic attribution.

In this work, we leverage real-time DNS responses gathered by the clients' default DNS resolvers.
Having these DNS responses and correlating them with network-layer headers, we are able to translate CDN-hosted domains to the actual services they belong to. 
We design a correlation system for this purpose and deploy it at a large European ISP.
With our system, we can correlate an average of 81.7\% of the traffic with the corresponding services, without any loss on our live data streams.
Our correlation results also show that 0.5\% of the daily traffic contains malformatted, spamming, or phishing domain names. 
Moreover, ISPs can correlate the results with their BGP information to find more details about the origin and destination of the traffic.
We plan to publish our correlation software for other researchers or network operators to use.
\end{abstract}

\setlength{\TPHorizModule}{\paperwidth}
\setlength{\TPVertModule}{\paperheight}
\TPMargin{5pt}

\begin{textblock}{0.8}(0.1,0.02)
    \noindent
    \footnotesize
    If you cite this paper, please use the CoNEXT '22 reference:
Aniss Maghsoudlou, Oliver Gasser, Ingmar Poese, and Anja Feldmann.
2022. FlowDNS: Correlating Netflow and DNS Streams at Scale. In \textit{The
18th International Conference on emerging Networking EXperiments and
Technologies (CoNEXT ’22), December 6--9, 2022, Roma, Italy}. ACM, New
York, NY, USA, 9 pages. \url{https://doi.org/10.1145/3555050.3569135}
\end{textblock}

\maketitle
\section{Introduction} \label{section:introduction}
ISPs need to know where their traffic is coming from originally not only to provide their customers with better quality, but also to better plan their network infrastructure and collaborations.

To this end, most of the service providers gather network-layer statistics of their traffic using different protocols, e.g. Netflow \cite{cisco-Netflow}, IPFIX \cite{ipfix}, etc. which usually include source and destination addresses, and traffic volume.%
\footnote{\new{Throughout this work, we refer to network flow information as Netflow, which is a standardized protocol to collect IP network traffic \cite{rfc3954}.}}
Network-layer headers do not contain the domain name of the services they belong to.
To complicate this even more, Over-The-Top (OTT) services are nowadays adopting multi-CDN approaches \cite{regalado2014}, making the inference of the service merely on the IP address nearly impossible.
Therefore, either DNS records or the application layer information are needed.
Nowadays, the application layer information is oftentimes encrypted \cite{felt2017} and therefore, not visible to the service providers.

DNS is one of the core services to map domain name to the IP address \cite{rfc1034,rfc1035}, and can be used to find the original source of the service. 
There have been several studies using machine learning techniques for domain name recognition, all of which use passive DNS records \cite{fastflux,li2020,bilge2011,bao2019,perdisci2020}.
Inspecting the traffic on a large European ISP, we observe that more than 85\% of the traffic is originated by CDNs.
These CDNs might use one domain name for several services in different locations/times and IP addresses could also be re-used \cite{rama2020}.
Therefore, IP address to domain name mappings change frequently in CDN-hosted domain names \cite{pad2020,rama2020}, and DNS records used for such correlation should not be outdated.
\new{Thus, capturing the DNS records collected from user requests is the most suitable way of mapping domain name to IP address. }

\new{Previous studies use software-defined networking to correlate DNS responses with web traffic either in an external controller \cite{netassay2014} or in the data plane \cite{meta42021,alsabeh2022}. However, all these approaches introduce parsing limitations, e.g. domain names length limit, and also ignore encrypted DNS packets.
This work, however, does not enforce any limitation on the DNS records, and is not affected by DNS encryption. Unlike the previous studies, we do not aim at direct policy enforcement and therefore, do not require any modification to the existing network architecture. 
We instead propose a system that can run on any machine receiving a flow of DNS records and network flows.}

Using DNS records from the same source as the traffic gives the domain name recognition more certainty.
In the meanwhile, using the same sources of DNS and Netflow translates to a higher processing load since both sources need to be processed synchronously either in a real-time fashion or offline.
In case the processing is to be done offline, the timestamps need to be taken into account and the two sources of data, namely Netflow and DNS records, need to be correlated in the window where the DNS record is still valid, i.e. TTL > 0.
Although research has shown that in some scenarios, longer TTLs can reduce latency \cite{moura2019}, our experiments show that 99\% of the record have TTLs of less than two hours.
Monitoring TTLs for every single record while correlating induces higher memory usage and lower correlation rates.
Multi-level caching in DNS resolvers makes this even more complicated \cite{randall2020}.

Correlating two live sources of data, each carrying thousands or millions of records per second, also requiring keeping some of the records for later use, is not trivial. 
Doing so with the standard database queries for an hour of data, takes tens of hours, making this correlation impossible to be done in real-time.
Therefore, we propose \FlowDNS{}, a system to correlate Network-layer headers and DNS records in real-time.
Our work consists of three main contributions:

\begin{itemize}[leftmargin=*]
\item We design, build, and deploy a system for real-time DNS-Netflow Correlation called \FlowDNS{} in a large European ISP. In \Cref{section:methodology}, we go over the system's building blocks, and in \Cref{section:evaluation}, we show that we can correlate 81.7\% of the data.
\item Using the correlated data in our deployment of \FlowDNS{}, we identify the traffic using malformatted, spam and phishing domains in \Cref{section:use-cases}. We observe that 0.5\% of the daily traffic volume uses either malformatted or spam/phish domain names.
\item Finally, we formulate our learned lessons in building a real-time DNS-Netflow Correlation system in \Cref{section:lessons}.
\end{itemize}

\section{Data Overview} \label{section:data-overview}
We use flow data and DNS traffic from a Large European ISP.
\new{The flow data is in Netflow format, and
we receive both Netflow and DNS traffic as live streams:}
 
\begin{itemize}[leftmargin=*]
\item \textbf{DNS streams}: A set of DNS \textit{cache misses} gathered from different customer resolvers. 
    For load-balancing purposes, the data is already divided into 2 different streams, carrying ~75K DNS records per second on average collectively.
    Each record in a DNS stream contains:
    \textit{timestamp,..., [name; rtype; ttl; answer] <0,n>}
\item \textbf{Netflow streams}: A set of Netflow records captured at the network ingress interfaces. 
    For load-balancing purposes, the data is already divided into 26 different streams. 
    These streams input ~1M Netflow records per second on average.
    Each record in a Netflow stream contains:
    \textit{..., srcIP, dstIP, ..., timestamp, packet, bytes}
\end{itemize}

We also deploy our system on a smaller European ISP with one DNS stream carrying 115K DNS records per second and two Netflow streams with 138K Netflow records per second.

Each of the above-mentioned streams has an internal buffer to be used in case the reading speed is less than their actual rate.
If that buffer overflows, the streams start to drop data. 
Throughout this paper, wherever we mention \textit{loss on the streams}, we mean that the buffers are overflown and start to drop. 
Therefore, the goal is to keep the buffer usage stable to avoid any loss.

Reading from multiple streams requiring shared memory access, and keeping the DNS records in memory to be quickly accessible makes this correlation challenging in terms of memory and CPU usage. 
To overcome these challenges, we leverage different techniques, details of which are explained in \Cref{section:methodology}.
In accordance with the data provider agreement, we refrain from reporting the exact values of the traffic, and all the traffic volume data throughout the paper is normalized.

\section{Methodology} \label{section:methodology}
The goal is to categorize source of the traffic by their service in near real-time, e.g. to understand what fraction of the traffic is originated from Netflix, Amazon Prime, Google, etc.
To realize this, we look for the IP address of the Netflow records in the \textit{answer} section of the A/AAAA DNS records to find the \textit{name} it corresponds to.
Then looking at the CNAME records, we search for that \textit{name} to find the corresponding CNAME.
The results from this correlation are then correlated with BGP information to find the information such as source/destination ASes for each service.
For the sake of brevity, in this work, we only focus on DNS-Netflow correlation.
\new{We note that the system is not bound to NetFlow data and can be adapted to use other data formats containing IP addresses and timestamps in a configuration file.}
\new{\subsection{Overview}\label{subsection:meth-overview}}

\begin{figure}[t]
  \begin{center}
  \includegraphics[width=0.9\linewidth]{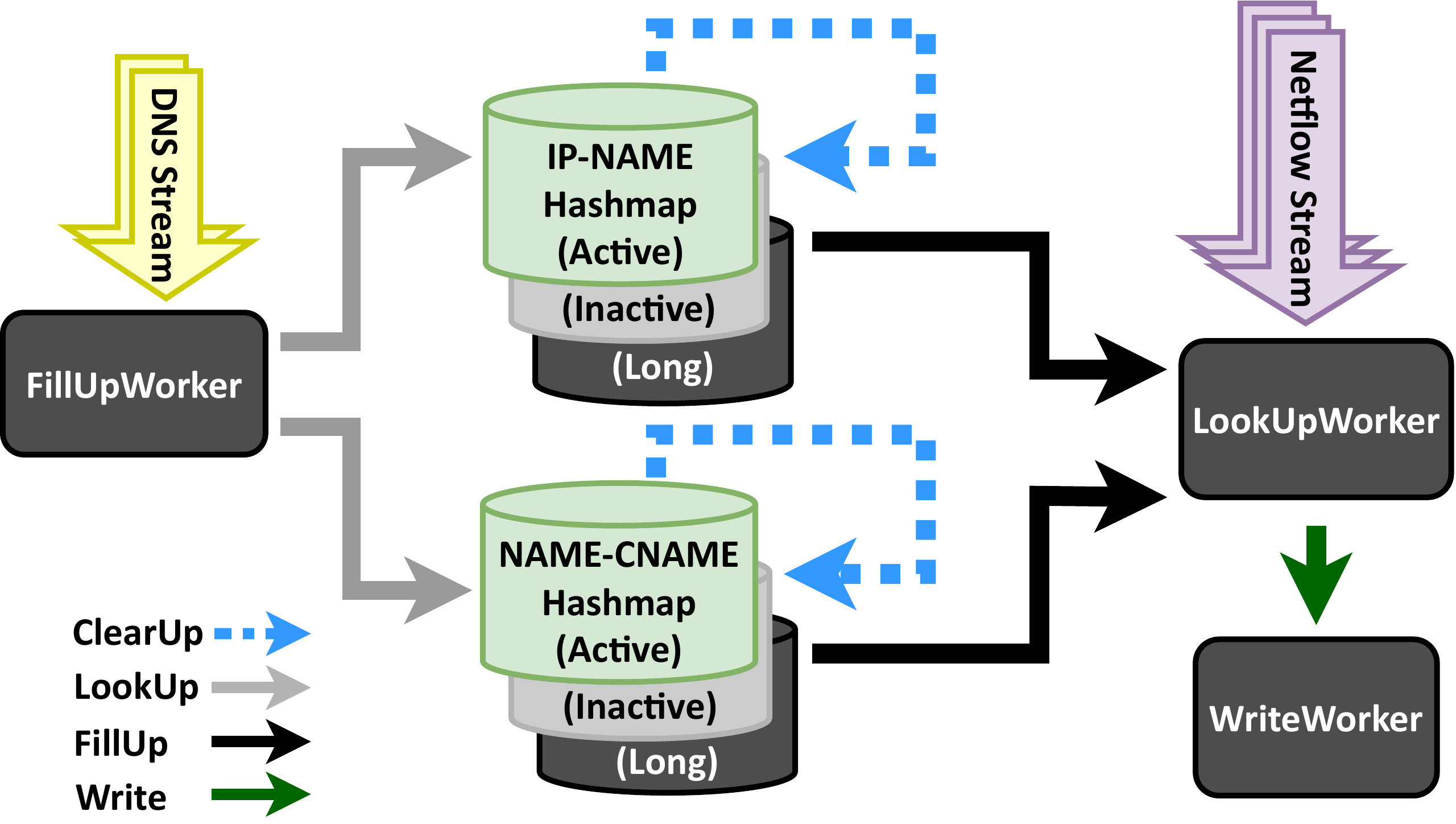}
  \end{center}
  \caption{\FlowDNS correlation architecture.}
  \Description{Parallel Processing of DNS and Netflow Records}
  \label{fig:correlation-architecture}
\end{figure}

To perform the DNS-Netflow correlation, as \Cref{fig:correlation-architecture} shows, the DNS streams are received by \textit{FillUp} workers. 
\new{Multiple FillUp workers are allocated to each DNS stream to enable parallel processing of each shard of the DNS stream.}
These workers analyze the DNS records and fill up a shared internal storage with the DNS records. 
At the same time, the Netflow streams are received by the \textit{LookUp} workers. These workers look for the source of the traffic in the shared internal storage. 
\new{Again, we assign multiple LookUp workers to each Netflow stream.}
In our work, we are interested in analyzing the source of the traffic, hence we use the source IP address. 
Nonetheless, destination address or both source and destination addresses can be used with minor modifications. 
Then the result of this lookup is written onto the disk by the \textit{Write} workers.
\new{Each worker has an input and output queue which enables the communication between workers.
It is important to avoid that too many workers write to the same queue, as this contention causes a decrease in performance.} 
Since multiple instances of the LookUp workers will try to simultaneously access a shared data structure where DNS records are kept, we split the DNS data and distribute them to different splits to then isolate each split as much as possible. \new{In \Cref{section:evaluation}, we discuss whether this further splitting is needed.}
In \Cref{subsection:dns-proc} and \Cref{subsection:netflow-proc}, we go through the steps starting from reading the streams, to fill up the internal storage, and then look up and write.

We cannot strictly apply DNS records TTL and expire them after the TTL has passed, since there are multiple levels of caching and each might apply a different tolerance for expiring the records. 
\new{Moreover, applying the actual TTLs on the DNS records requires regular iterations over all DNS data to check their TTL expiration.
This degrades the performance dramatically and increases the memory usage (cf. \Cref{sub-section:exact-ttls}).}
On the other hand, we cannot keep the DNS records forever due to memory constraints. 
Therefore, we need to clear up the storage.
We observe that 99\% of the A/AAAA and CNAME records have a TTL smaller than 3600 and 7200 seconds, respectively (Ref. \Cref{sub-section:dns-ttls}).
Therefore, we clear up the A/AAAA records every 3600 seconds, and CNAME records every 7200 seconds.
However, since clearing the whole storage will remove all the states, we perform buffer rotation before clear-up.
\new{The internal storage where we keep the DNS records is a hashmap with the DNS \textit{answer} section as key, and the \textit{query name} as value. For implementing these hashmaps, we use the \textit{concurrent-map} module in Go \cite{concurrent-map}, which allows for high-performance concurrent reads and writes by sharding the map.}

\new{We add the DNS records in a primary hashmap, \ie the \textit{active} hashmap. 
After a certain amount of time has passed, we copy the contents of the active hashmaps into a secondary storage, \ie the \textit{inactive} hashmap, and clear up the active hashmap.}
In the next clear-up round, the current contents of the inactive hashmap will be over-written by the new contents.
Active hashmaps are actively updated with the newly arrived DNS records, while inactive hashmaps are only updated when the active hashmaps are cleared.
A very small fraction of the DNS records has TTLs longer than the clear-up interval. 
Therefore, in case a DNS record's TTL is larger than a certain threshold, we put it into specific hashmaps which are never cleared or are cleared much less frequently, namely, long hashmaps.
Otherwise, it is stored in the active hashmap.
From now on, we call the active/inactive hashmaps for A/AAAA and CNAME records IP-NAME\textsubscript{active/inactive}, and NAME-CNAME\textsubscript{active/inactive} respectively. 
See \Cref{sub-section:appendix-parameters} for an overview of the parameters and hashmaps used in \FlowDNS.

\subsection{DNS Processing}\label{subsection:dns-proc}
This part of \FlowDNS takes in DNS streams and fills up an internal shared storage with the DNS records. 
These records will then be accessed in the Netflow Processing. 
\begin{enumerate}[wide, labelwidth=!, labelindent=0pt]
\item DNS Streams are received by separate threads.
\item The DNS records go through a filter to check if they are valid DNS responses.
\item Valid DNS responses are added to a queue, namely FillUp Queue, to be then processed \new{each by several} FillUp workers. 
We need this queue to facilitate the synchronous execution of different workers.
\item \label{step:label-dns} Each FillUp worker picks a DNS response from the FillUp Queue and if it is an A/AAAA record, labels it based on the IP address. This label will be used as a hashmap index later on.
\item The FillUp worker then puts the DNS response in the shared hashmaps. 
In all our hashmaps, the key is the answer section, and the value is the query.
We leverage two kinds of hashmaps: 
\begin{itemize}
\item \textbf{IP-NAME hashmap:} Maps the answer section in an A/AAAA response, i.e. the IP address, to the queried domain name.
\new{We divide these hashmaps into several splits. 
We empirically find that 10 splits are suitable for our scenario. This can change for any deployment depending on the traffic volume.}
If, in Step \ref{step:label-dns}, the IP for an A/AAAA response gets the label \textit{{n}}, $0\leq n<10$, it goes to \textit{IP-NAME\textsubscript{n}}. 
\item \textbf{NAME-CNAME hashmap:} Maps the answer section, i.e. the domain name, to the queried canonical name for CNAME records.
\end{itemize}
\item The FillUp worker keeps track of the timestamp in each DNS record. 
If \textit{AClearUpInterval} seconds has passed, it copies the contents of \textit{IP-NAME\textsubscript{active}} to \textit{IP-NAME\textsubscript{inactive}} and clears \textit{IP-NAME\textsubscript{active}}.
If \textit{CClearUpInterval} seconds is passed, it copies the contents of \textit{NAME-CNAME\textsubscript{active}} to \textit{NAME-CNAME\textsubscript{inactive}} and clears \textit{NAME-CNAME\textsubscript{active}}.
\end{enumerate}
See \Cref{sub-section:appendix-dns} for the pseudocode of the above algorithm.

\begin{comment}
\begin{figure}[t]
  \begin{center}
    \includegraphics[width=\linewidth]{figures/cdf-ttl}
  \end{center}
  \caption{Cumulative distribution of TTLs for DNS records over a day.}
\label{fig:cdf-ttl-plot}
\end{figure}
\end{comment}

\subsection{Netflow Processing}\label{subsection:netflow-proc}
In parallel with the DNS processing, this part of \FlowDNS{} takes in the Netflow streams, takes the source IP address, and looks for this IP address in the internal shared storage to find the corresponding domain name.
It then writes the results into the output files.
\begin{enumerate}[wide, labelwidth=!, labelindent=0pt]
\item Netflow Streams are received by separate threads.
\item The Netflow records go through a filter to check if they are valid Netflow records.
\item The valid Netflow records are added to the LookUp Queue to be processed \new{each by several LookUp} workers.
\item \label{step:firstlookup} Each LookUp worker picks a Netflow record from the LookUp Queue and labels it based on the \textit{srcIP} field.
\item The LookUp worker looks for the \textit{srcIP} in the \textit{IP-NAME\textsubscript{active n}} hashmap, if the label from Step \Cref{step:firstlookup} is n.
If nothing is found, it looks into the Inactive hashmap, and next into the long hashmap.
If a \textit{Name} is found, the search continues onto the next step. 
Otherwise, the search finishes here for that \textit{srcIP} (\textit{result} = NULL).
\item The search will be continued in \textit{NAME-CNAME\textsubscript{active n}} to find the CNAME for that \textit{Name}.
If a \textit{CName} is found, the search continues to the next step.
Otherwise, the search continues in the Inactive and then the long hashmap. 
If nothing is found, the search finishes here for that \textit{Name} (\textit{result = Name}).
\item The search in the NAME-CNAME map continues until no further CNAME is found or a pre-defined loop limit is reached (\textit{result = CName}). Our experiments show that a loop limit of 6 is sufficient for more than 99\% of the records (ref. \Cref{sub-section:appendix-cnamechain}). If the \textit{result} is found with more than one look-up in NAME-CNAME maps, we add it to \textit{NAME-CNAME\textsubscript{active}} for later use.
\item The \textit{result} along with the original Netflow is then passed to the Write Queue to be written in the output file by WriteWorkers.
\end{enumerate}
See \Cref{sub-section:appendix-netflow} for the pseudocode.
We publish the code for \FlowDNS \cite{flowdns} for future researchers or network operators.
\section{Evaluation}\label{section:evaluation}

\begin{figure}[t!]
  \begin{center}
   \begin{subfigure}{0.4\textwidth}
    \includegraphics[width=\linewidth]{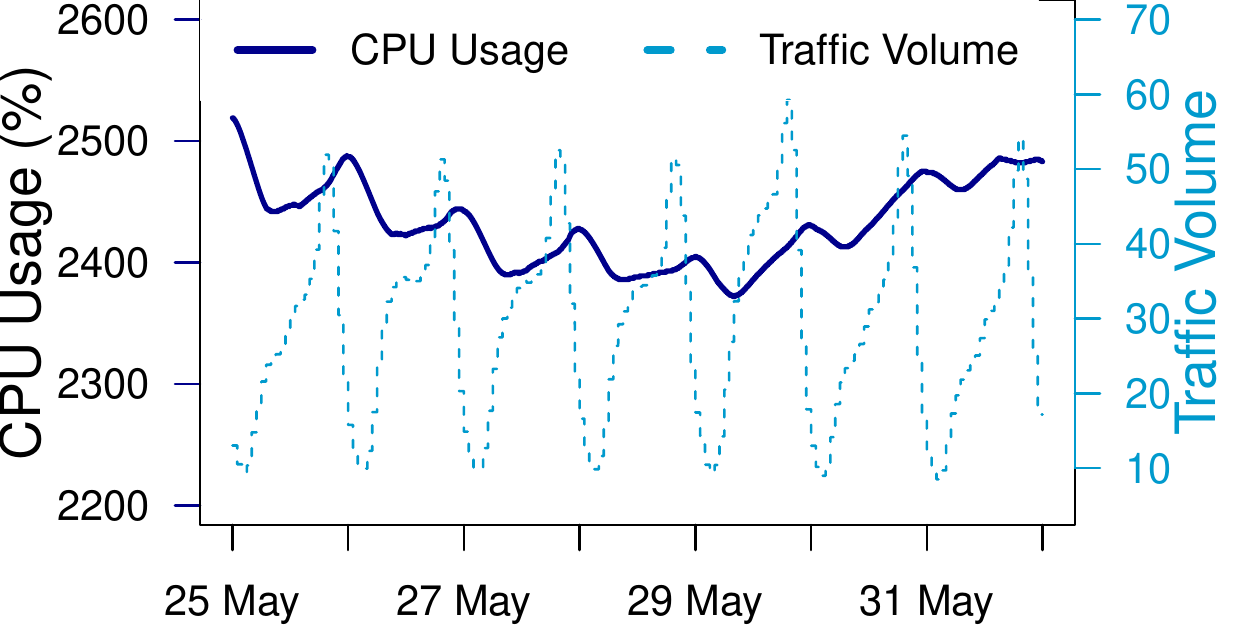}
    \caption{CPU usage} \label{fig:cpumain}
   \end{subfigure}
   \hspace*{\fill}
   \begin{subfigure}{0.4\textwidth}
    \includegraphics[width=\linewidth]{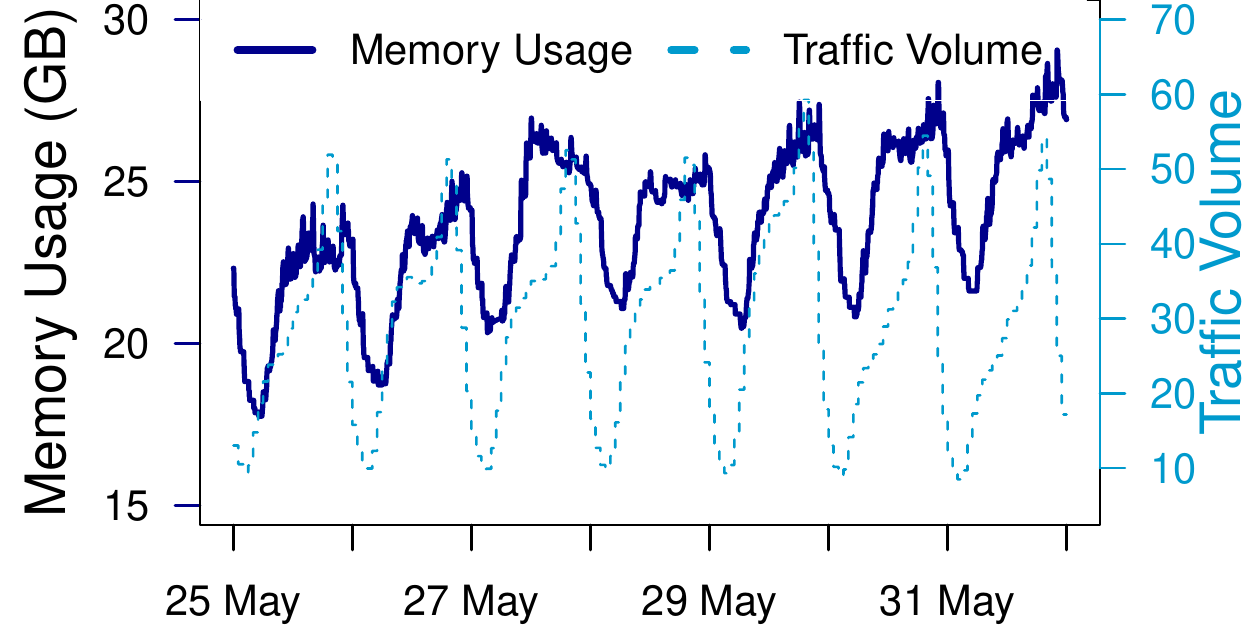}
    \caption{Memory usage} \label{fig:memmain}
   \end{subfigure}
   \hspace*{\fill}
  \end{center}
  \caption{CPU and memory usage for \textit{Main} benchmark over a week.}
  \label{fig:mainscript}

\end{figure}

In this section, we evaluate the matching accuracy and performance metrics for \FlowDNS implemented in Go.
First, we analyze the final version of \FlowDNS on a full week of traffic of a large European ISP.
Second, we selectively remove implementation features from \FlowDNS on a one-day traffic capture to understand their importance by evaluating the effect on matching accuracy, CPU usage, and memory consumption.
\new{We evaluate all the benchmarks on an Ubuntu 18.04.5 LTS machine with 128 cores and 756 GB RAM.}
\Cref{fig:mainscript} shows the CPU and memory usage of \FlowDNS over one week when deployed at a large European ISP.
\new{In both plots, the right axis shows the traffic volume to compare the CPU/memory usage patterns with the traffic load.}
For all three metrics\new{---traffic volume, memory usage, and CPU usage---}we can clearly identify diurnal patterns, with daily peaks in the evening period, a low time during night hours, and an increase during the day.
Note that we normalize the traffic volume in the right Y-axis.
\new{We show CPU usage as percentages, \ie every 100\% means 1 fully utilized CPU core.}
The CPU usage is around 2500\% which means roughly 25 CPUs are used. Memory usage also oscillates between 15 GB and 30 GB. 
In addition to the large European ISP, we also deploy \FlowDNS on a smaller network.
On the smaller network, we observe average memory usage of 6 GB, and CPU usage of around 300\%, both following a diurnal pattern.
The ratio of CPU usage and number of flows remains the same in both deployments.
The memory usage, however, is affected by both number of DNS records and number of parallel threads. This results in lower memory usage in the smaller network.
On both deployments, the results are written to disk by a maximum delay of 45 seconds, and without any significant loss, i.e. 0.01\% loss, on the data stream buffers.
The ratio of correlated traffic to the total traffic, i.e. the correlation rate, is 81.7\% on average for both deployments.
\new{We cannot correlate 18.3\% of the traffic since  
(1) the coverage of our DNS data is only 95\%, as discussed later in this section, and 
(2) not all the traffic is DNS-related, i.e. not all traffic has the destination IP address obtained through a DNS query.}

Now, we remove the techniques used in the fully featured version of \FlowDNS once at a time, introducing four new benchmarks:
\begin{itemize}[leftmargin=*]
\item \textit{No Split}: The hashmaps are not divided into several splits. 
\item \textit{No Clear-Up}: The hashmaps are kept in memory forever.
\item \textit{No Rotation}: The hashmaps are cleared, but no buffer rotation takes place and no Inactive hashmap exists.
\item \textit{No Long Hashmaps}: The hashmaps are cleared up, and buffer rotation takes place, but records with large TTLs are also written in the Active hashmaps, instead of the Long hashmaps.
\end{itemize}

\begin{figure}[t!]
  \begin{center}
   \begin{subfigure}{0.4\textwidth}
    \includegraphics[width=\linewidth]{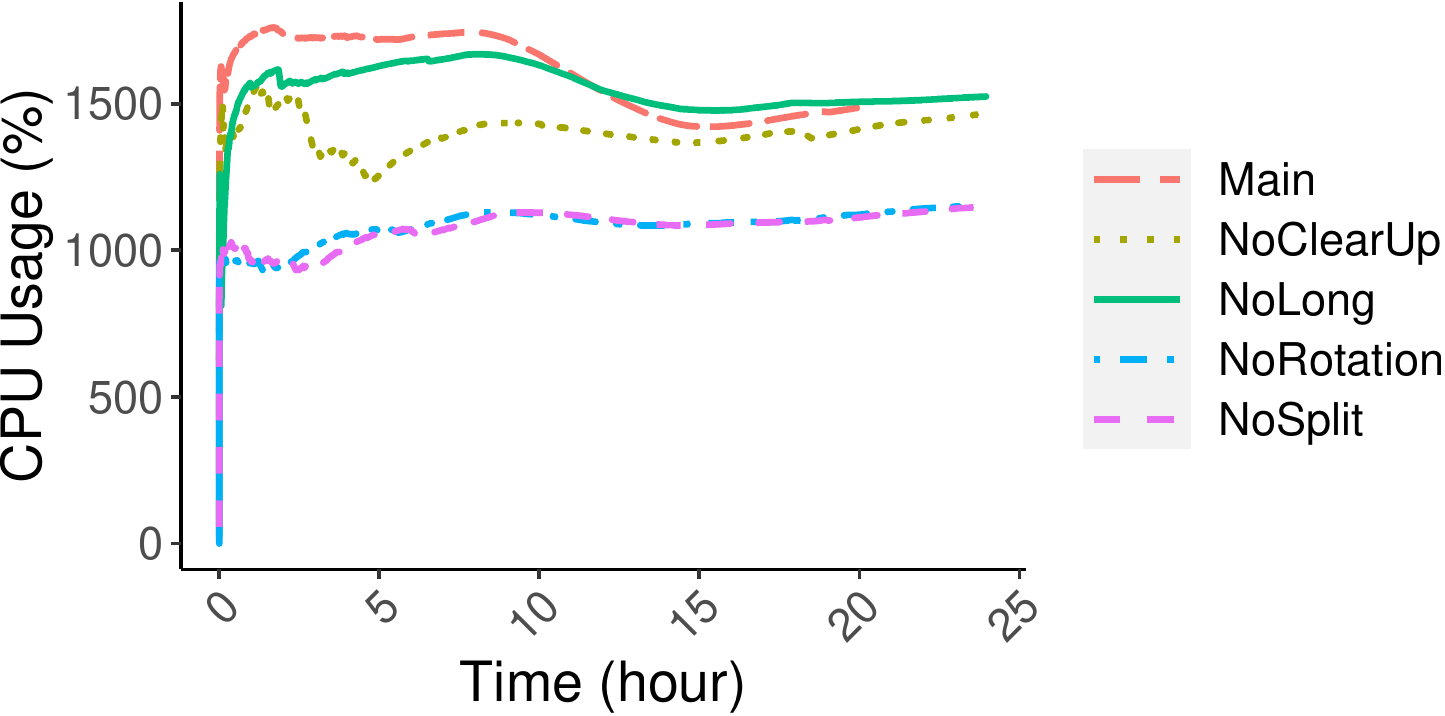}
    \caption{CPU usage} \label{fig:cpuvars}
   \end{subfigure}
   \hspace*{\fill}
   \begin{subfigure}{0.4\textwidth}
    \includegraphics[width=\linewidth]{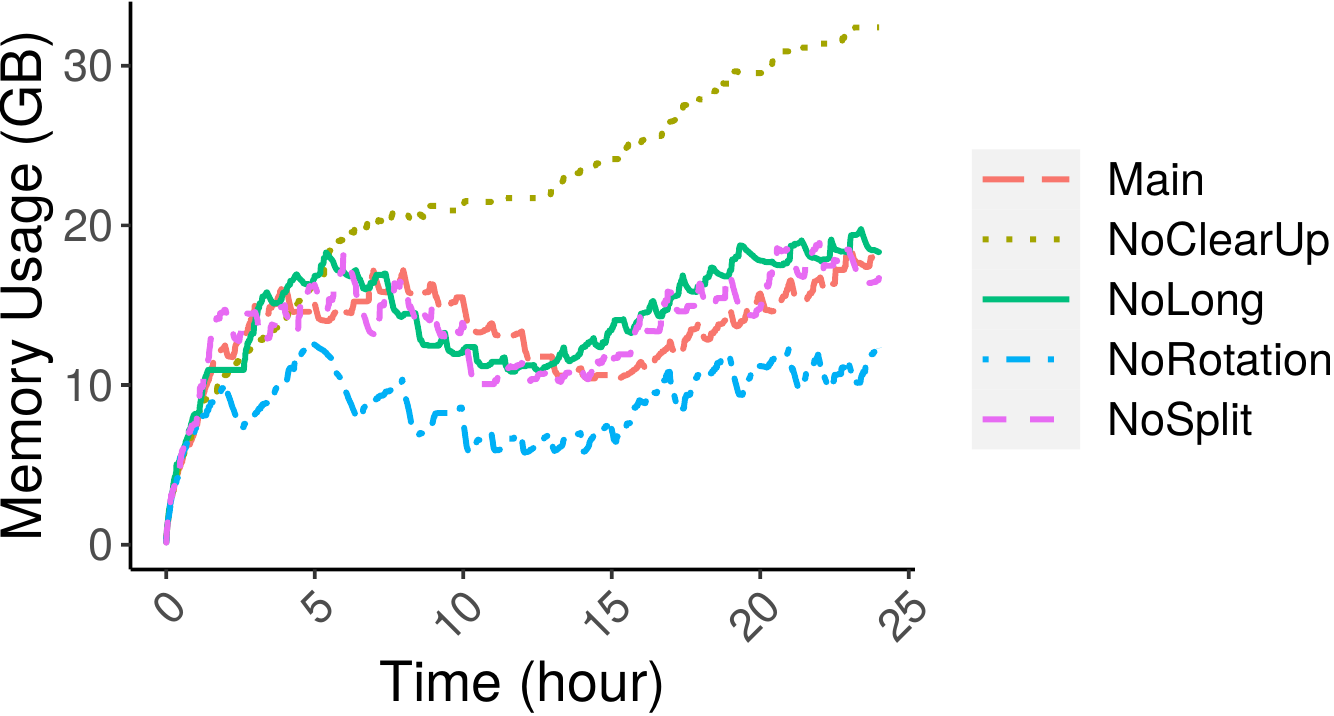}
    \caption{Memory usage} \label{fig:memvars}
   \end{subfigure}
   \hspace*{\fill}
  \end{center}
  \caption{CPU and memory usage for different variants over a day.}
  \label{fig:variants}
\end{figure}

\Cref{fig:variants} shows CPU and Memory usage for the above benchmarks. As expected, memory usage for \textit{No Clear-Up} grows steadily over the day and can easily hit the memory limit. The mean correlation rate for this benchmark is 82.8\%.
The \textit{No Rotation} benchmark uses much less memory compared to other benchmarks since it does not keep a copy of the original contents before clear-up. However, the average correlation rate for this benchmark is 79.5\%.
The \textit{No Long Hashmaps} save neither a significant amount of memory nor CPU, yet, with a correlation rate of 81.1\%, reduce the correlation rate by 0.6\% compared to the \text{Main} benchmark. Therefore, the Long Hashmaps help keep those DNS records from being cleared up without much cost.
The \textit{No Split} neither improves nor degrades the memory usage but decreases the CPU usage significantly. 
This could be due to the reduced effort to access separate hashmaps simultaneously. 
The average correlation rate is also 81.7\%.
\new{However, this should not be interpreted as if sharding the data is not helpful at all. In contrast, as we have used data sharding already both in our hashmaps, and in our job queues, explained in \Cref{subsection:meth-overview}. The fact that this feature does not help as much as the others only shows that no further splitting is needed in our case.}
\todoog{nosplit interpretation changed:should we remove the nosplit benchmark and also splitting mechanism explanation in methodology since it's apparently not needed?}
See \Cref{sub-section:appendix-correlationrate} for correlation rates per hour for all the benchmarks.

As we have seen with these four benchmarks, all implemented features in \FlowDNS, except for IP-splitting, help increase the correlation rate while keeping the CPU and memory usage low.

\new{
\textbf{Coverage}.
\FlowDNS receives DNS cache misses gathered from the clients' default ISP resolvers. 
This data is sent from the ISP resolvers to our collectors via TCP.
Therefore, even if the client uses DNS encryption while still using the default ISP resolver, the results from \FlowDNS are not affected.
However, if the clients use a resolver other than the default ISP resolver, e.g. a public DNS resolver (\eg Cloudflare's 1.1.1.1, Google Public DNS, or Quad9), the DNS record is not received and therefore, \FlowDNS can not correlate the Netflow traffic for those clients.
To understand the number of DNS records we lose due to clients using public DNS resolvers, we analyze a sample 1-hour Netflow data and filter DNS and DoT traffic, i.e. ports 53 and 853.
Then, using a public DNS resolvers list \cite{public-dns-resolvers} and comparing it with our sample, we observe that 1 out of every 20 DNS packets is sent to a public DNS resolver. 
Therefore, the coverage of our DNS data is 95\%.
}

\new{
\textbf{Accuracy}.
Earlier in \Cref{section:evaluation}, we report that the correlation rate for \FlowDNS, i.e. the number of bytes that could be correlated with \textit{any} service/domain name compared to the total traffic volume in bytes, is 81.7\% on average.
However, this metric does not show whether the correlated service is the service actually used by the clients.
Since we cannot access the actual domain names used by the clients, there is no ground truth against which we can compare our results.
Nevertheless, we can estimate the accuracy of \FlowDNS, by pinpointing the scenarios that could result in an incorrect service correlation and estimate their impact on the system's accuracy.
}

\new{
In \FlowDNS, we keep the DNS data in a hashmap with the IP address as the key and domain names as values. Therefore, by design, observing multiple IP addresses for one domain name in the DNS data does not affect the accuracy of \FlowDNS.
However, observing multiple domain names for one IP address can affect the accuracy.
In case a second domain name is observed with the same IP, i.e. the same key, the existing (first) domain name is overwritten by the second domain name, which in turn decreases the accuracy of our system.
To confirm this, we design a small-scale accuracy analysis using generated traffic data. 
We browse two different websites and capture the traffic.
Then, we extract the DNS packets from the captured traffic and feed them to \FlowDNS as the DNS stream.
We then create Netflow records from all traffic packets and feed them to \FlowDNS as the Netflow stream. 
Finally, observing the correlated domain names and comparing them to the actual scenario, we find whether the system has correlated correctly. 
We consider two scenarios for this experiment: 
(1) Two websites with different domain names and different IP addresses.
(2) Two websites with different domain names, using the same IP address. 
In the first scenario, we observe that all the traffic is correlated correctly, while in the second scenario, all the traffic is correlated to the second domain name.
In other words, we had an accuracy of 100\% and 50\% in the first and second scenarios, respectively.
}%

\new{
To estimate the impact of such mislabelling events, we analyze the domain name distribution per IP address.
To this end, we analyze a 300-second sample of DNS records since as \Cref{fig:cdf-ttl-plot} shows, more than 70\% of the DNS records have TTL < 300 seconds. We observe that 88\% of IP addresses are mapped to only a single domain name, as shown in \Cref{fig:numname}. We also did the analysis with a 1-hour sample and observed similar results.
Therefore, we expect our results to be accurate for 88\% of IP addresses in our flow data.
}
\begin{comment}
\begin{figure}
  \begin{center}
    \includegraphics[width=0.5\linewidth]{figures/cdf-numname}
  \caption{Cumulative distribution of number of domain names per IP address.}
  \label{fig:numname}
 \end{center}
\end{figure}
\end{comment}
 %
\section{Use Cases}\label{section:use-cases}

\FlowDNS{} helps ISPs to better plan their networks, while providing the opportunity to analyze the traffic originated by malicious IDN homographs and spam domain names. 
There have been several studies on detecting malicious or unwanted domain names \cite{pala2020,afzal2021,yadav2010,xiaoqing2019}, detecting IDN homographs \cite{yazdani2020,zhu2020,suzuki2019}, and also analyzing domain classification services \cite{vallina2020}.
However, to the best of our knowledge, there is no work measuring the traffic going to/originated by these domains.
In this section, we illustrate three example use cases of \FlowDNS, measuring the traffic from malicious or malformed domain names.

\begin{figure}
  \begin{center}
   \begin{subfigure}{0.49\textwidth}
    \includegraphics[width=\linewidth]{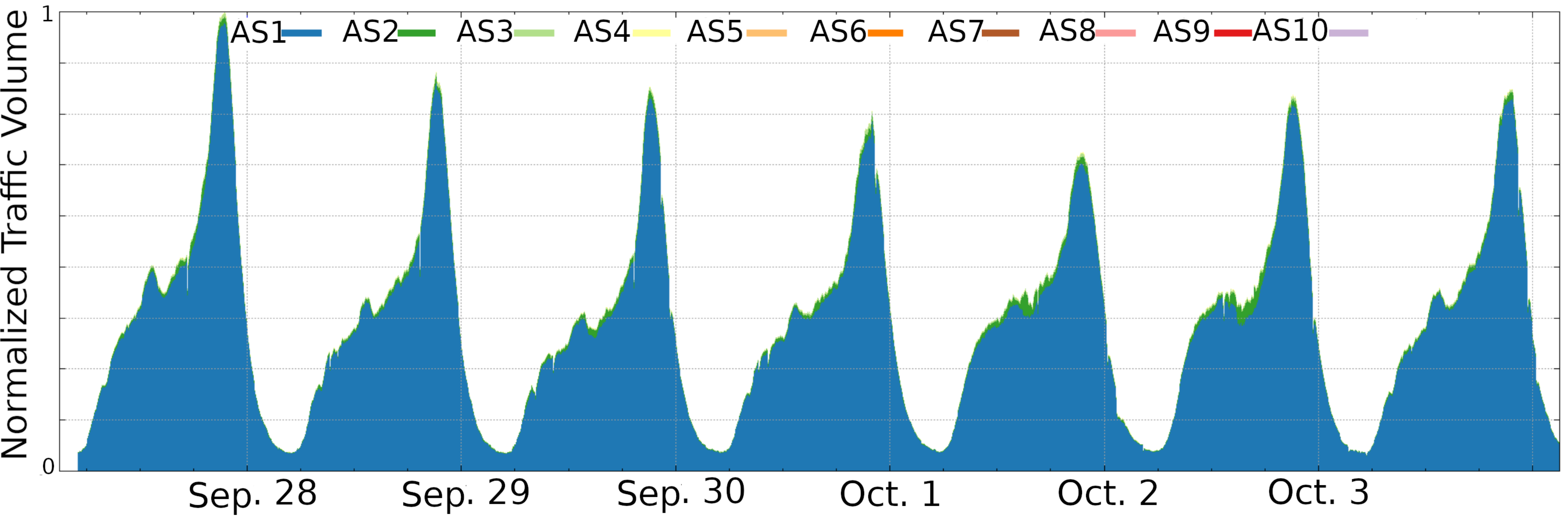}
    \caption{Streaming service \textit{S1}}
    \label{fig:v-report}
   \end{subfigure}
   \hspace*{\fill}
   \begin{subfigure}{0.49\textwidth}
    \includegraphics[width=\linewidth]{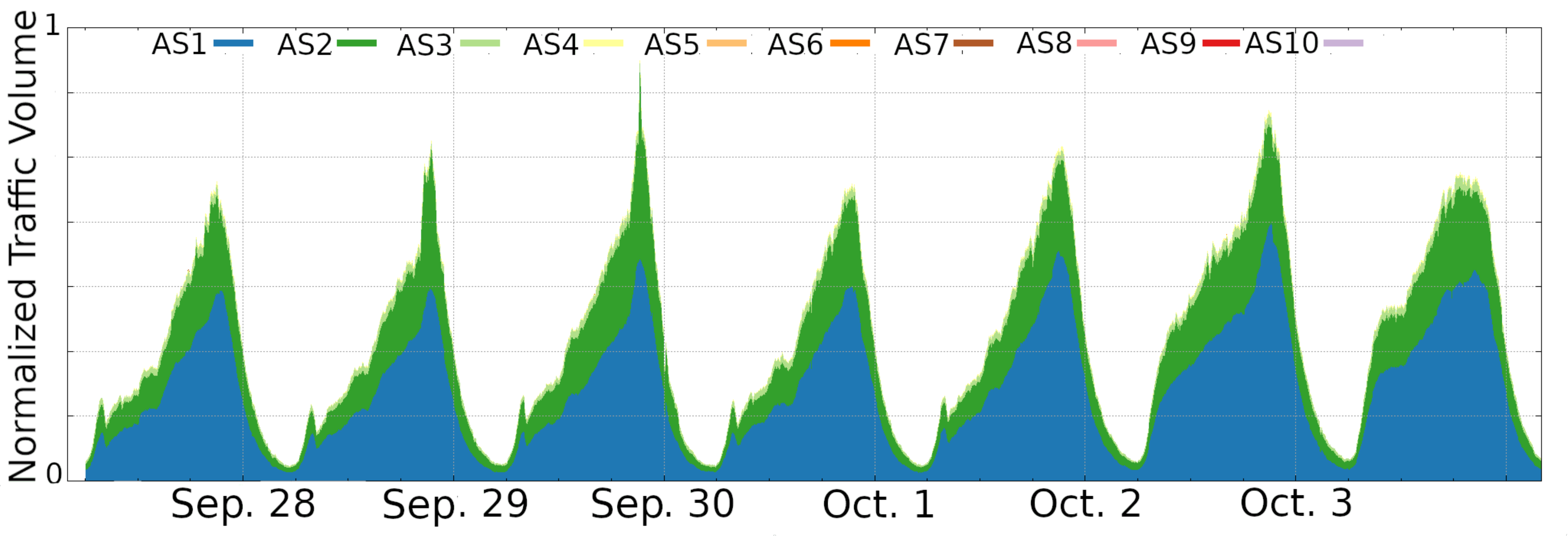}
    \caption{Streaming service \textit{S2}}
    \label{fig:d-report}
   \end{subfigure}
   \hspace*{\fill}
  \caption{\new{Cumulative traffic volume for streaming service \textit{S1} and \textit{S2} per source AS.}}
  \label{fig:reports}
 \end{center}
\end{figure}

For all the following use cases, we use the correlated traffic for over a day in a large European ISP, including 39M unique domain names, and analyze the traffic originated by these domain names. 
\new{
\textbf{Network Provisioning and Planning}.
\FlowDNS is already deployed in a large European ISP and a smaller European ISP. The output from \FlowDNS is then correlated with BGP data, e.g. source AS, destination AS, hand-over AS, etc., to gain more knowledge about the path the traffic of a specific service takes. 
\Cref{fig:reports} shows the contribution of different source ASes to the traffic volume of streaming services S1 and S2 over a week at the ISP.
As \Cref{fig:v-report} shows, the traffic corresponding to the streaming service S1 is originated mostly from only one AS, while the streaming service S2 is originated mainly by two ASes as shown in \Cref{fig:d-report}. Note that AS numbers in two figures do not represent the same ASes necessarily.  
In \Cref{fig:reports} we observe a diurnal pattern with slight differences between the two services.
Knowing the source and intermediate ASes serving a specific service helps ISPs to negotiate with content providers over using ISP's resources instead of a third-party CDN. Also, in case of a broken peering link, it helps find the fallback paths, if they will be overloaded, and which services are effected. 
}

%
\begin{comment}
\textbf{IDN Homographs}.
Internationalized Domain Names (IDNs) are domain names that contain at least one special non-Latin character. 
Sometimes, malicious parties abuse non-Latin characters that look similar to Latin characters to deceive users and lead them to their own domain names \cite{liu2018}. 
We look at the domain names in our correlated traffic to see if any traffic goes to IDN homographs. 

To detect actual homographs attacks, we capture all the correlated traffic for a day and filter IDNs. 
Then, using the Unicode confusable characters list \cite{confusables}, we generate all the permutations of those IDNs to find out whether any well-known domain name is generated out of them.
Then, we cross-check the generated permutated domain names with the domain names in our 1-day traffic capture. 
We do not find any match between the two, probably because the Unicode character list is not comprehensive enough and misses some visually confusable mappings. 
Therefore, we manually check all IDNs from the actual traffic, to identify IDN homographs.
We specifically find homographs of \textit{google.com}, i.e. googlê.com, góogle.de, góogle.de, etc., and a Cyrillic homograph of \textit{apple.com}.

The amount of traffic going to these IDN homographs is minimal, i.e. only hundreds of megabytes are originated by these homographs in a day.
This shows that \FlowDNS can be a valuable tool for network operators to identify potentially ongoing phishing campaigns of their users early on.
%
\end{comment}

\begin{figure}
  \begin{center}
    \includegraphics[width=\linewidth]{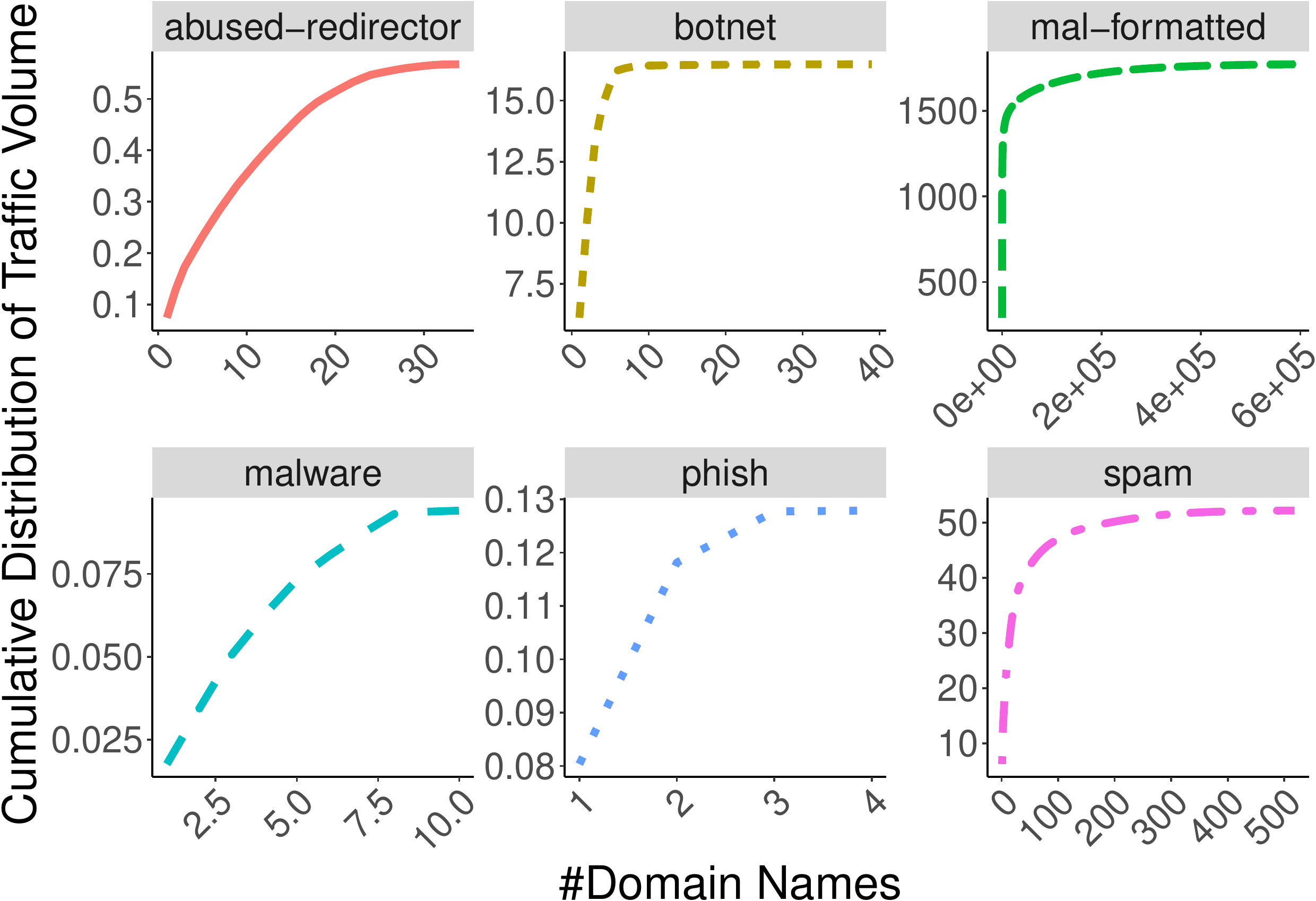}
  \caption{Cumulative distribution of the traffic volume per number of domain names.}
  \label{fig:use-cases}
 \end{center}
\end{figure}

%
\textbf{Spam Domains}.
Using our 1-day traffic capture, we check the correlated domain names with the Spamhaus DBL (Domain Block List) \cite{spamhaus-dbl} to see if any spamming, phishing or otherwise suspicious domains are generating any traffic.
To avoid bandwidth limitations on Spamhaus DBL, we sample all the domain names once every hour, giving ca. 1M domain names, out of which 612 are classified as suspicious by the Spamhaus DBL. These include 512 \textit{spam/generic bad reputation domains}, 41 \textit{botnet C\&C domains}, 34 \textit{abused spammed redirector domains}, 11 \textit{malware domains}, and 3 \textit{phishing domains}. 
Collectively, these suspicious domain names originate multiple terabytes of traffic.
\new{\Cref{fig:use-cases} shows a cumulative distribution of traffic volume per number of domain names for each of the above categories. In other words, it shows how many domain names contribute to what fraction of the traffic volume. } 
As can be seen, a significant amount of traffic comes from spam and botnet domains, while only a limited number of domain names account for a large fraction of the traffic.

Malicious websites usually change their domain names rapidly to avoid being detected. Therefore, spam detection datasets such as Spamhaus DBL have an expiry date for their labels, i.e. if checked after the expiry date, they will no longer exist in the dataset and therefore be labeled as benign. \FlowDNS{} allows for real-time checking of the domain names with such datasets.

\textbf{Invalid Domain Names}.
RFC 1035 stipulates specifications of DNS domain names \cite{rfc1035}.
In the following analysis, we focus on three rules to which valid domain names must adhere:
\begin{itemize}[leftmargin=*]
\item The total length of the domain name is 255 bytes or less.
\item Each label in the domain name is limited to 63 bytes. 
\item Each label starts with a letter, ends with a letter or digit, and the interior characters are limited to letters, digits, and hyphens.
\end{itemize}
The word \textit{label} refers to each part of the domain name separated by dots, i.e. if the domain is \textit{A.B.C.com}, labels are: \textit{A},\textit{B},\textit{C}, and \textit{com}.
In our 1-day traffic capture, we observe that %
666k domain names violate at least one of the above-mentioned rules. \Cref{fig:use-cases} shows that almost all the traffic comes from a very limited number of domain names, and the amount of traffic originated by such domain names is quite significant. Note that the traffic volume is normalized.
The most common conflict with the above-mentioned rules is disallowed interior characters. 
The most common disallowed character found in 87\% of the malformatted domains is the underscore character, i.e. "\textbf{\_}".
Finally, we investigate the overlap of invalid domain names with domains in the spam category and find that only four malformatted domains also appear in the spam category. %
\new{To understand how clients treat these malformed domains, we investigate whether traffic is being exchanged for these domains.
We observe that 2.7\% of the clients which receive traffic from malformed domains, send traffic back to 23.6\% of these malformed domains.
This bi-directional traffic accounts for 1.9\% of the packets, mostly related to non-web port numbers, e.g. OpenVPN and Kerberos.
All other packets are originated by malformed domains and receive no answer.}
\section{Lessons Learned}\label{section:lessons}
During the design of \FlowDNS, we learned the following lessons:

\begin{itemize}[leftmargin=*]
\item 
When following the CNAME chain, we had to limit the chain length to 6 due to performance reasons.
    In our experiments, we observed that less than 1\% of CNAME chains are longer than 6. 
\item Splitting the data into several shards allows for higher parallelism, while consuming higher CPU for the same amount of data. \new{Therefore, it is important to keep an eye on this trade-off.}
\item Buffer rotation, i.e. copying the data once before clearing it, helps to increase correlation percentage without substantial CPU or memory usage in the long run.
    Therefore, it provides a good trade-off between resource utilization and correlation percentage.
\item \new{Expiring DNS records using their exact TTLs induces an unnecessary contention over the shared memory 
making the loss rate reach over 90\%.}
    Using rotating buffers with a common expiry time instead of the exact value helps in gaining the same correlation rate compared to keeping the DNS records forever, with no loss and is much more resource-efficient.
\end{itemize}

We hope that these lessons will prove useful for fellow network application developers and researchers alike.
\section{Conclusion}\label{section:conclusion}
Inferring the services behind a certain traffic flow is not possible merely by looking at the IP addresses due to the prevalent deployment of CDNs. 
In this work, we presented \FlowDNS, a system to correlate DNS and Netflow streams in real-time. 
We used several techniques such as splitting the data, rotating buffers, and specific hashmaps to keep track of longer-living DNS records.
We evaluated each of these techniques and confirmed the usefulness of each.
Then, using \FlowDNS, we analyzed the domain names with known datasets to detect malicious domains and observed that a substantial amount of traffic is originated by these domain names.
Moreover, we checked the adherence of those domain names to standardization rules and observed that 1.7\% of all the domain names violate them.
The traffic originated by such domains accounts for 0.5\% of the daily traffic.
Finally, we plan to make \FlowDNS available to fellow researchers and network operators.

\bibliographystyle{ACM-Reference-Format}
\bibliography{paper}

\appendix
\section{Appendix}

\subsection{Parameters and In-memory Storage}\label{sub-section:appendix-parameters}
\Cref{tab:parameters} shows an overview of the parameters and names of the in-memory storage that we use in \FlowDNS. Note that \textit{$ 0 \leqq n < NUM\_SPLIT $ }.

\begin{table}[t]
\resizebox{\columnwidth}{!}{
    \begin{tabular}{ p{2.9cm}@{\hskip 0.3cm}p{6cm}@{\hskip 0.3cm} }
        \toprule
        \textbf{Parameter Name}           & \textbf{Description}\\ %
        \midrule
        AClearUpInterval        & Time in seconds after which the \textit{IP-NAME} hashmap is cleared.\\
        CClearUpInterval    & Time in seconds after which the \textit{NAME-CNAME} hashmap is cleared.\\
        NUM\_SPLIT              & Number of splits for each IP-NAME hashmap. \\
        \midrule
        \textbf{Storage Name}    & \textbf{Description} \\
        \midrule
        IP-NAME\textsubscript{Active n}   & Hashmap for DNS records with TTL < AClearUpInterval and label n.\\
        IP-NAME\textsubscript{Inactive n}   & Hashmap where the contents of IP-NAME\textsubscript{Active n} are copied to every AClearUpInterval seconds\\
        IP-NAME\textsubscript{Long n}         & Hashmap for the new DNS records with TTL >= AClearUpInterval and label n.\\
        NAME-CNAME\textsubscript{Active}   & Hashmap for the new CNAME responses with TTL < CClearUpInterval.\\
        NAME-CNAME\textsubscript{Inactive}   & Hashmap where the contents of NAME-CNAME\textsubscript{Active} are copied to every CClearUpInterval seconds\\
        NAME-CNAME\textsubscript{Long}         & Hashmap for the new CNAME responses with TTL >= CClearUpInterval.\\
        \bottomrule
    \end{tabular}\textsc{}
}
\caption{Overview of Parameters and Storage Names.}
\label{tab:parameters}
\end{table}

\subsection{DNS Processing Pseudocode}\label{sub-section:appendix-dns}
Algorithm \ref{alg:fillUpWorker} shows an overview of the fillUpWorker thread which we mentioned in \Cref{subsection:dns-proc}. This function takes the DNS records and fills in the hashmap.

\subsection{Netflow Processing Pseudocode}\label{sub-section:appendix-netflow}
Algorithm \ref{alg:lookUpWorker} shows an overview of the lookUpWorker thread which reads the netflow records, looks them up in the Active, Inactive, and Long hashmaps and finds the results.

\subsection{CNAME Chain Length Distribution}\label{sub-section:appendix-cnamechain}
\begin{figure}
  \begin{center}
    \includegraphics[width=\linewidth]{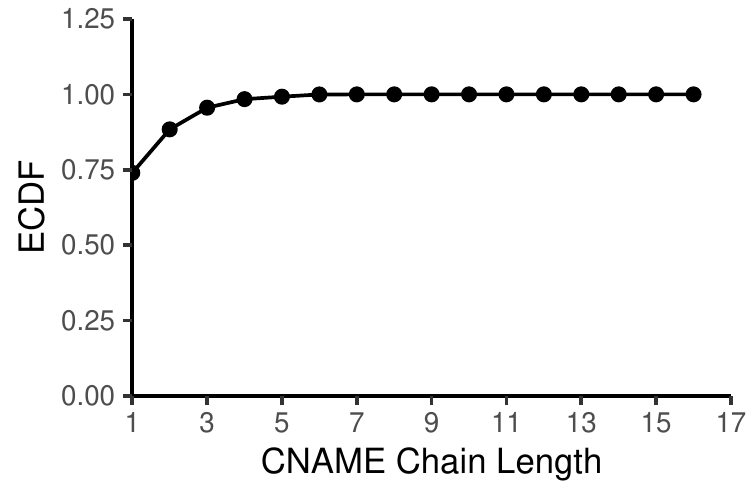}
  \end{center}
  \caption{Cumulative distribution of CNAME chain length over a day.}
\label{fig:cdf-cnamechain}
\end{figure}

CNAME look-up can sometimes include multiple consequent look-ups with one CNAME mapping to another more than once. We studied the CNAME chain length, and as shown in \Cref{fig:cdf-cnamechain}, we observed that more than 99\% of the DNS records can be mapped with a chain of 6 look-ups. \new{Therefore, we limit the number of CNAME chain look-ups to 6 in \FlowDNS.}

\subsection{Correlation Rate}\label{sub-section:appendix-correlationrate}
\begin{figure}
  \begin{center}
    \includegraphics[width=\linewidth]{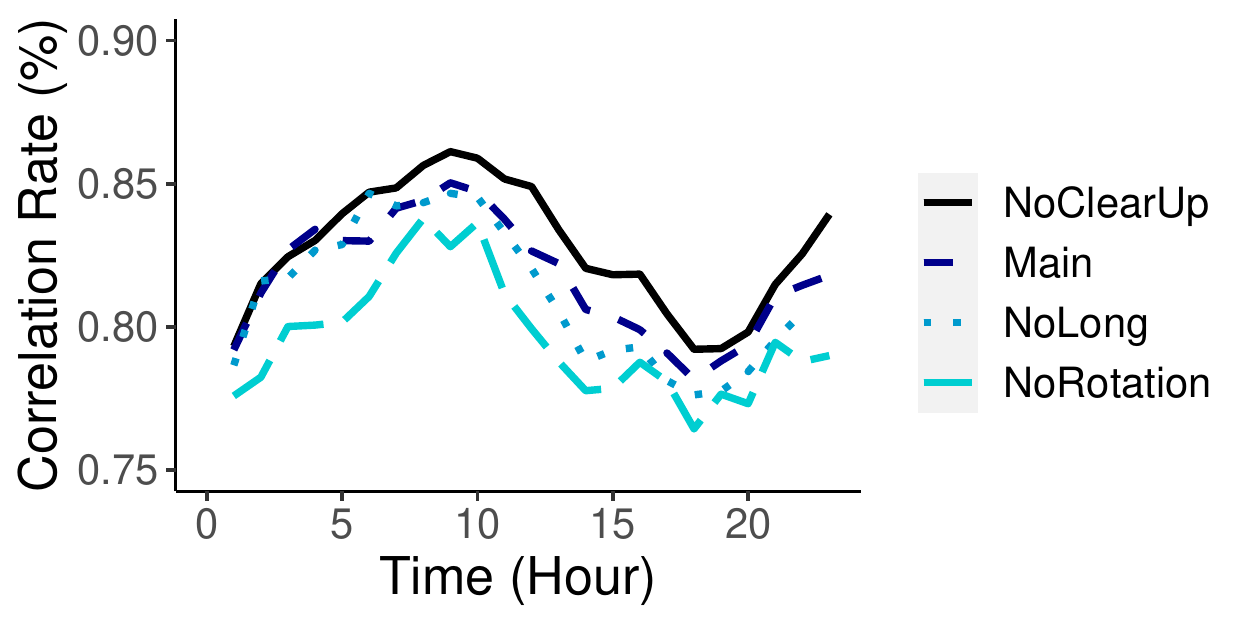}
  \end{center}
  \caption{Correlation rate for benchmark variants.}
\label{fig:correlationrate}
\end{figure}

The correlation rate, i.e. the ratio between correlated traffic and total traffic is illustrated in \Cref{fig:correlationrate} for different benchmark variants. 
The \textit{No Split} benchmark excluded from the plot since it has a complete overlap with the \textit{Main} benchmark.
The top two variants in terms of correlation rate are \textit{Main} and \textit{NoClearUp}. The \textit{NoClearUp} performs unacceptable in terms of memory usage. The lowest correlation rate belongs to \textit{NoRotation}, which shows the importance of buffer rotation in \FlowDNS.

\subsection{\new{DNS Records' TTLs}}\label{sub-section:dns-ttls}
\begin{figure}[t]
  \begin{center}
    \includegraphics[width=\linewidth]{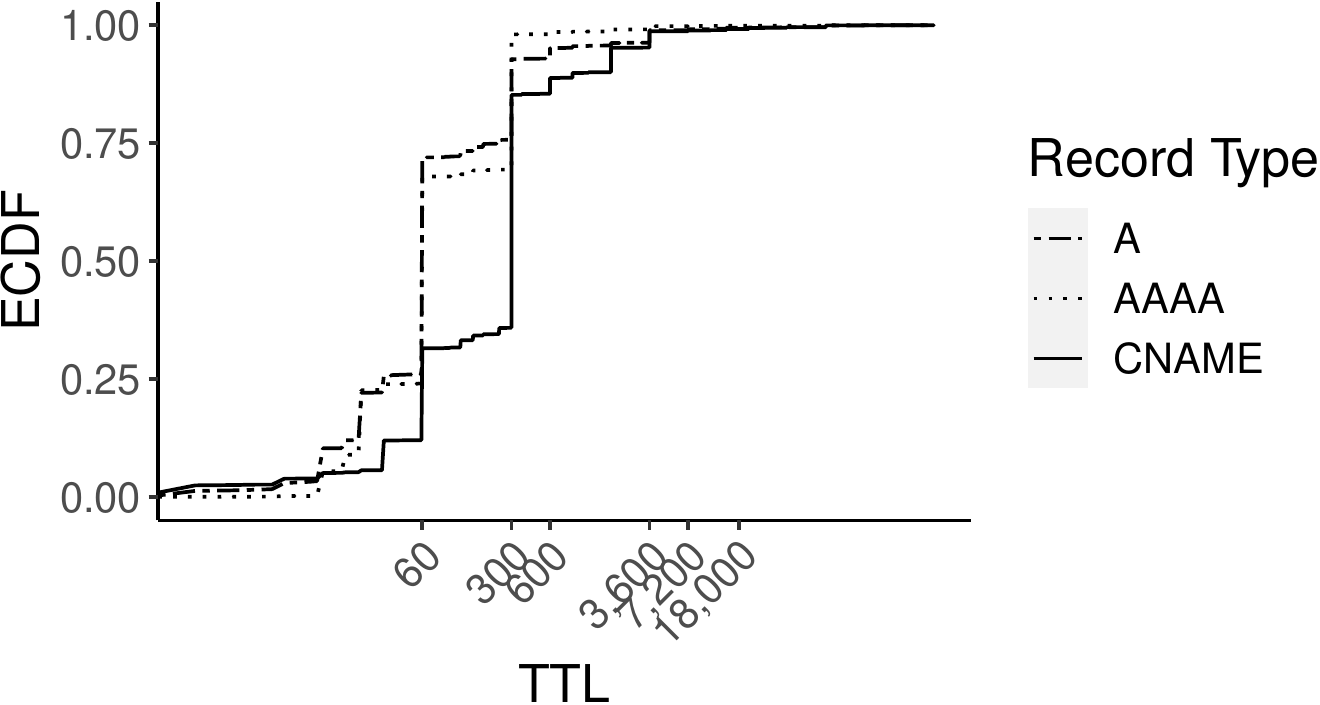}
  \end{center}
  \caption{Cumulative distribution of TTLs for DNS records over a day.}
\label{fig:cdf-ttl-plot}
\end{figure}

\new{To find out the correct number for the clear-up intervals, namely \textit{CClearUpInterval} and \textit{AClearUpInterval}, we investigate the TTLs for the DNS records. We look at the DNS records' TTLs over a day at a large Europoean ISP, and find out that 99\% of the A/AAAA and CNAME records have TTL smaller that 3600 and 7200 seconds respectively, shown in \Cref{fig:cdf-ttl-plot}. Therefore, in \FlowDNS, we set the clear-up variables as follows: \[ CClearUpInterval = 7200 \] \[AClearUpInterval = 3600 \]}

\subsection{\new{Number of Domain Names per IP address}}\label{sub-section:ip-numnames}
\begin{figure}
  \begin{center}
    \includegraphics[width=\linewidth]{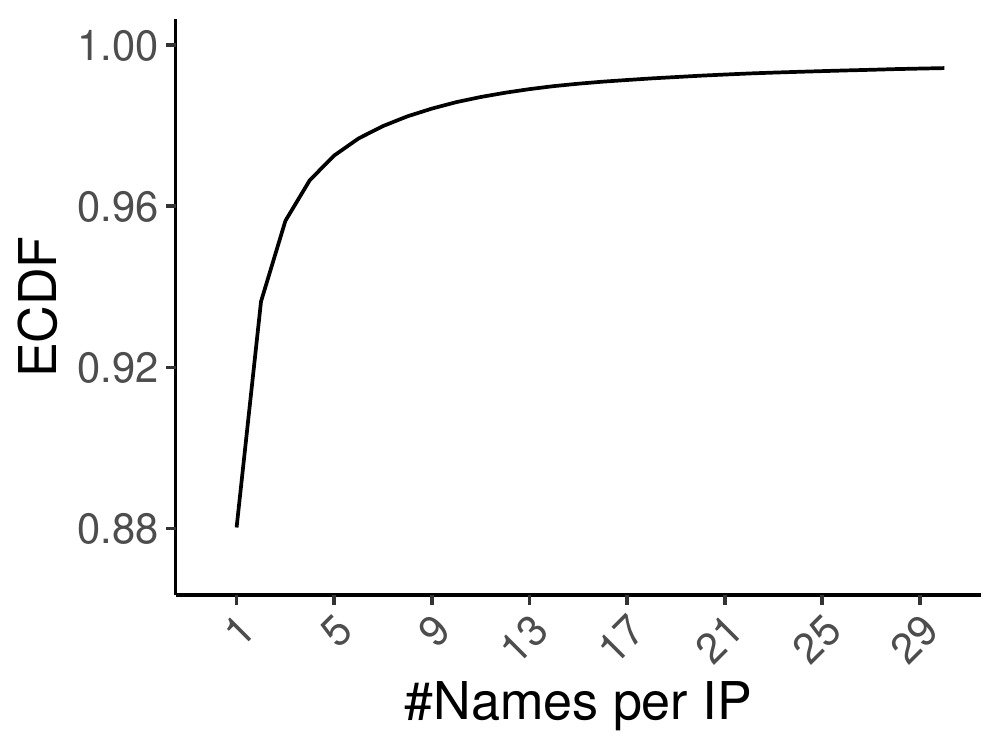}
  \caption{\new{Cumulative distribution of number of domain names per IP address.}}
  \label{fig:numname}
 \end{center}
\end{figure}

\new{We look at a 300-second period of DNS records to investigate the number of domain names that map to the same IP address leading to a mislabelling event in \FlowDNS. \Cref{fig:numname} shows the cumulative distribution of number of domain names per IP address. We observe that ~88\% of the DNS records only map to one domain name in 300 seconds. Note that we choose 300 seconds since this is the TTL for ~70\% of our DNS records. We also analyze this in a 1-hour sample of DNS data and observe similar results.}
\new{Additionally, we analyze the number of IP addresses per domain name in a 300-second period of DNS records. We observe that 35\% of the domain names map to more than one IP address. We also analyze this in a 1-hour sample of DNS records and observe similar results. Note that observing multiple IP addresses per domain name, which is a significantly more probable event compared to multiple names for one IP address,does not effect the accuracy of \FlowDNS.}

\subsection{\new{Applying the Exact TTLs}}\label{sub-section:exact-ttls}
\new{We try applying the exact TTLs from the DNS records on our correlation, meaning we correlate the IP from a DNS record with the source IP from the Netflow record only if the DNS record's TTL plus its timestamp is less than the timestamp from the Netflow record which we consider current timestamp. In other words:
\[ TTL\textsubscript{dns} + Timestamp \textsubscript{dns} < Timestamp \textsubscript{netflow} \]
We also run a regular process to clear-up the expired DNS records, when the above-mentioned condition does not hold.
We run this on the same sources of data, meaning DNS and Netflow streams at the large European ISP.
We observe that the internal buffers of all the streams start to overload from the very first minutes of running the above-mentioned system, with the loss rate of over 90\% for both Netflow and DNS streams.
We observe that the memory usage reaches up to 45 GB memory usage after only 1 hour of running the system.
When we compare this to the results from \FlowDNS in \Cref{fig:memmain}, we see that the memory usage is doubled although only 10\% of the data is received at the system and others are lost.
This could be due to the regular clear-up process not being fast enough to clear-up all the expired TTLs as the hashmaps grow, while at the same time, the contention to access the shared memory is so high that the performance degrades dramatically.
}

\begin{algorithm}[t]
\caption{DNS Read and Fill-up Overview}
\label{alg:fillUpWorker}
\SetKwProg{fillUpWorker}{Function \emph{fillUpWorker}}{}{end}
\fillUpWorker{DNSRecord d}{
    $n$ = label($d$)\;
     \uIf{ $d$.rtype is A/AAAA }{
        \If{ d.ts - lastAClearUpTs >= 3600 }{
            IpName.Inactive = IpName.Active\;
            IpName.Active = \{\}\;
            lastAClearUpTs = d.ts\;
        }

        \uIf{ $d$.ttl <= 3600}{
            IpName.Active[$n$][$d$.answer] = $d$.query\;
        }
        \Else{
            IpName.Long[$n$][$d$.answer] = $d$.query\;
        }            
      }
    \uElse{
        \If{ d.ts - lastCClearUpTs >= 7200 }{
            NameCname.Inactive = NameCname.Active\;
            NameCname.Active = \{\}\;
            lastCClearUpTs = d.ts\;
        }

        \uIf{ $d$.ttl <= 7200}{
            NameCname.Active[$n$][$d$.answer] = $d$.query\;
        }
        \Else{
            NameCname.Long[$n$][$d$.answer] = $d$.query\;
        }
    } 
}
\end{algorithm}

\begin{algorithm}
\caption{Netflow Read and Look-up Overview}
\label{alg:lookUpWorker}
\SetKwProg{lookUpWorker}{Function \emph{lookUpWorker}}{}{}
\lookUpWorker{NetflowRecord $nf$}{
    n = label(nf)\;
    Name = deepLookUp(nf.srcIP, IpNameObj[n])\;
    loopCount = 0\;
    \uIf {Name != Null}{
        results = append(results, Name)\;
        Cname = deepLookUp(Name, NameCnameObj[n])\;
        \While {Cname != Null \textbf{and} loopCount <= 6}{
            results = append(results, Cname)\;
            loopCount ++\;
        }
    }
    \KwRet results\;
}
\DontPrintSemicolon
\;
\PrintSemicolon
\SetKwProg{deepLookUp}{Def \emph{deepLookUp}}{}{}
\deepLookUp{NetflowRecord nf, MapObj hm}{
    Name = NULL\;
    \uIf { nf.srcIP in hm.Active}{
        Name = hm.Active[nf.srcIP]\;
    }
    \uElseIf{ nf.srcIP in hm.Inactive}{
        Name = hm.Inactive[nf.srcIP]\;
    }
    \uElseIf{ nf.srcIP in hm.Long}{
        Name = hm.Long[nf.srcIP]\;
    }
    \KwRet Name\;
}
\end{algorithm}

\end{document}